\documentclass[a4paper,11pt]{article}
\pdfoutput=1

\usepackage{ifxetex}
\ifxetex
  \usepackage{fontspec} % For XeLaTeX or LuaLaTeX
  \usepackage{xunicode} % For XeLaTeX or LuaLaTeX
\else
  \usepackage[utf8x]{inputenc} % For PDFLaTeX
\fi

\usepackage{cite}
\usepackage{amsmath, amssymb}
\usepackage{graphicx}
\usepackage{braket}
\usepackage{fullpage}
\usepackage{import}
\usepackage[hyperref]{xcolor}

\usepackage{hyperref}
\hypersetup{
    hypertexnames=false,
    colorlinks=true,
    anchorcolor=blue,
    citecolor=blue,
    filecolor=blue,
    linkcolor=blue,
    menucolor=blue,
    urlcolor=blue,
    pdfborder = {0 0 0},
    pdfstartview={FitH},
    pdfauthor={Andrius Štikonas},
    pdftitle={},
    bookmarksnumbered=true,
    linktoc=all
}
\DeclareMathOperator{\tr}{tr}

\author{Andrius Štikonas}
\title{Scrambling time from local perturbations of the rotating BTZ black hole}

\numberwithin{equation}{section}

\begin{document}

\maketitle

\abstract{In this paper, we investigate the entanglement entropy of the rotating BTZ black hole perturbed by a massive back-reacting free falling particle. Then, mutual information between two finite intervals in two asymptotic regions of rotating BTZ is derived. It allows us to find the scrambling time, the time scale in which mutual information vanishes. We give a dual large $c$ CFT description in terms of a thermofield double state with different temperatures for left and right moving modes that is perturbed by a local operator. Exact matching between gravity and CFT results is obtained.}

\tableofcontents

\section{Introduction}

The behaviour of perturbed quantum thermal systems has been an active research area recently, in particular in the context of AdS/CFT correspondence \cite{Maldacena:1997re}. Entanglement entropy~\cite{vonNeumann:1927:TQG}, which is a measure quantifying how entangled two subsystems of a quantum system are, is a useful tool for studying the behaviour of quantum field theories. It is particularly useful in AdS/CFT correspondence where entanglement entropy has a simple geometric interpretation of Ryu-Takayanagi surface \cite{Ryu:2006bv, Ryu:2006ef}.

Over the last few years a lot of work has been done in investigating entanglement entropy and scrambling in perturbed systems \cite{0808.2096v1, Shenker:2013pqa, Shenker:2013yza, Roberts:2014isa, Reynolds:2146285, Leichenauer:2014nxa,Caputa:2014eta, PAM_doi:10.1007/JHEP08(2016)106,1412.6087v3, Caputa:2015waa,Caputa:2015qbk,Caputa:2017ixa,BenTov:2017kyf} both in the CFT context and in holography. Many of these papers investigate the chaotic behaviour of perturbed thermal system with the help of out of order (OTO) correlators of the form $\braket{V(0)W(t)V(0)W(t)}$ \cite{PAM_doi:10.1007/JHEP08(2016)106}. Holographically they are modelled by shock wave geometries introduced in \cite{Shenker:2013pqa} where BTZ black hole is perturbed at time $-t_\omega$ in the past with the in-falling null spherical shell, hence creating a discontinuity in the metric. The scrambling time if found by computing the smallest time $t_\omega$ such that mutual information between two subsystems $I_{A:B} = S_A + S_B - S_{A \cup B}$ vanishes. Note that mutual information gives an upper bound on the correlators \cite{0704.3906v2}
\begin{equation}
  I_{A:B} \geq \frac{\left(\braket{ \mathcal{O}_A \mathcal{O}_B} - \braket {\mathcal{O}_A } \braket{ \mathcal{O}_B }\right)^2}{2\|\mathcal{O}_A\|^2\| \mathcal{O}_B\|^2}.
\label{eq:bound}
\end{equation}
Therefore, the time scale in which mutual information becomes zero can be used to estimate when the two subsystems become uncorrelated. This approach is particularly useful in two dimensions where problem of calculating entanglement entropy is more analytically tractable. However, in some cases it is still possible to obtain results in higher dimensional cases. For example, back-reaction caused by shock-wave in rotating Kerr-Newman solution was found in \cite{BenTov:2017kyf}.

In our previous paper \cite{Caputa:2015waa} the scrambling time was computed in two dimensional large $c$ CFT in thermofield double (TFD) state. We also computed the same quantities for the dual holographic system which is described by \emph{static} BTZ black hole. We found exact matching between the bulk and boundary theories.

In this paper we will generalize the results of \cite{Caputa:2015waa} to the \emph{rotating} BTZ black hole.
In \autoref{s:CFT} we will consider a $\mathrm{CFT}$ in the TFD state with different temperatures for holomorphic and antiholomorphic modes which is the $\mathrm{AdS/CFT}$ dual of the rotating BTZ black hole. Then we perturb the CFT with a primary operator. The $\mathrm{CFT}$ part of the calculation is a reasonably straightforward generalization of the previous work but we try to keep it self-contained. The holographic dual of this system is a free falling massive\footnote{By massive we mean its mass is significant enough to cause back-reaction on the geometry.} particle in the rotating BTZ black hole which we discuss in \autoref{s:bulk}. We will use the fact that the BTZ black hole geometry locally looks like anti de-Sitter space to find the back-reacted geometry.

Both in  the field theory and gravitational calculations we find that the scrambling time is
\begin{equation}
t^\star_\omega=y + \frac{L}{2}-\frac{\beta_+}{2\pi}\log\left(\frac{\beta_+}{\pi\epsilon}\frac{\sin\pi \alpha_\psi}{\alpha_\psi}\right)+\frac{\beta_+}{2\pi}\log\left(4 \sinh\frac{\pi L}{\beta_+}\sinh\frac{\pi L}{\beta_-}\right),
\end{equation}
where $\alpha_\psi$ carries information about the conformal dimension of perturbation (mass of the particle) and $\epsilon$ is a UV cut-off for the energy of local excitation\footnote{$\epsilon$ should not be confused with UV cut-off for the field theory $\epsilon_\mathrm{UV}$ which is much smaller. Similarly in the gravitational theory one is initial position of the massive particle while the other becomes cut-off location for entanglement surface.}, $y$ is the distance from the point of perturbation to the region of entanglement and $L$ is its size.

\section{CFT results}
\label{s:CFT}

Let $\beta_\pm$ be the inverse temperatures of the CFTs (or inverse temperatures of inner and outer horizons in the gravitational theory),

\subsection{Thermofield Double State}

Consider two 2d CFTs with Hamiltonian $H$ and momentum $P$ on Hilbert spaces $\mathcal{H}_L$ and $\mathcal{H}_R$. We can then construct an entangled thermofield double state on $\mathcal{H}_L \otimes \mathcal{H}_R$ \cite{Maldacena:2001kr,Hartman:2013qma,PAM_doi:10.1007/JHEP11(2013)052,Hubeny:2007xt,PAM_doi:10.1007/JHEP01(2015)036}
\begin{equation}
 \ket{\Psi_\beta} = \frac{1}{\sqrt{\mathcal{Z} (\beta_-, \beta_+)}} \sum_n \exp\left(-\beta_+ E_{+,n}/2 - \beta_- E_{-,n}/2 \right) |n \rangle_L |n \rangle_R
\end{equation}
where $\ket{n}_L$ and $\ket{n}_R$ are simultaneous eigenstates of operators $H_{L,R} = \frac{H \pm P}{2}$ with eigenvalues $E_{\pm,n}$ and $\beta_\pm$ can be interpreted as the inverse temperatures of right and left moving modes. In the equation above $\mathcal{Z} (\beta_-, \beta_+)$ is the grand canonical partition function
\begin{equation}
 \mathcal{Z}(\beta_-, \beta_+) = \tr \exp \left( -\beta_+ H_L - \beta_- H_R \right).
\end{equation}
% Note that $\beta_\pm = \beta (1 + \Omega)$, where $\Omega$ is the chemical potential corresponding to the angular momentum.

\subsection{Setup}

Mutual information in the perturbed TFD state was discussed in \cite{1211.2887v2,Hartman:2013qma,PAM_doi:10.1007/JHEP11(2013)052}. In this section we will generalize the formalism developed in \cite{Caputa:2015waa} where perturbed TFD state was dual to the non-rotating eternal BTZ black hole. In this paper we will look at the rotating but not extremal BTZ black hole and its CFT dual, so $\beta_+$ and $\beta_-$ no longer have to be equal.

We define regions $A$ and $B$ to be intervals $[y, y + L]$ on the left and right boundaries respectively. Then mutual information is given by
\begin{equation}
 I_{A:B} = S_A + S_B - S_{A \cup B}.
\end{equation}

As in \cite{Caputa:2015waa} we consider TFD state excited with a primary operator $\psi(0, -t_\omega)$ in the CFT$_L$ at time $-t_\omega$ in the past. In order to calculate density matrix, we will have two copies of $\psi$ inserted at the same point. This can be avoided by spreading out perturbation on a scale $\epsilon$ which we choose to be much larger than ultraviolet cut-off $\epsilon_{\mathrm{UV}}$. Then time-evolved\footnote{We will use $t_-$ and $t_+$ to denote time coordinates in $\mathrm{CFT}_L$ and $\mathrm{CFT}_R$ respectively.} reduced density matrix is

 \begin{equation}
   \rho_L(t_-) = \mathcal{N} e^{-iH_L t_-} e^{-\epsilon H_L} \psi(0,-t_\omega) e^{2 \epsilon H_L + \beta_+ H_L} \psi^\dagger(0,-t_\omega) e^{-\epsilon H_L} e^{i H_L t_-}.
 \end{equation}
 
Density matrix can be expressed in a simpler form using the Euclidean time and complex coordinates
\begin{equation}
 \rho_L(x_1, \bar{x}_1, x_4, \bar{x}_4) = \mathcal{N} \psi(x_2, \bar{x}_2) e^{\beta_+ H_L} \psi^\dagger(x_1, \bar{x}_1),
\end{equation}
where operator insertion points are
\begin{align*}
 & x_1=-i\epsilon,\quad x_4=\hphantom{-}i\epsilon, \\
 & \bar{x}_1=\hphantom{-}i\epsilon,\quad \bar{x}_4=-i\epsilon.
\end{align*}
 
\begin{figure}[htb]
  \centering
  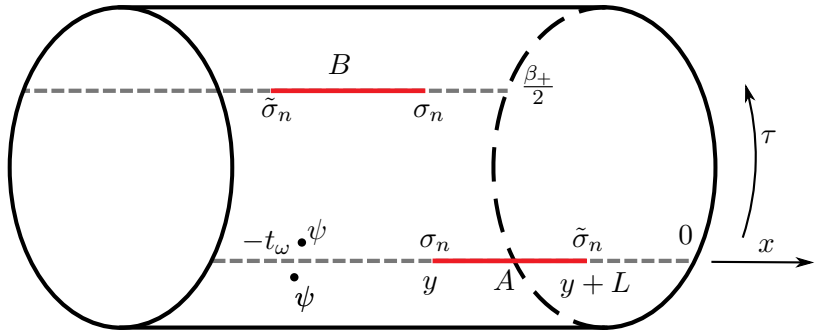
  \caption{Entanglement regions and operator insertion points on the thermal cylinder.}
  \label{fig:cylinder}
\end{figure}
% Is beta_+ better (since other coordinates are without bars, or beta_\pm
 
To calculate Rényi entropy we use twist operators that glue different copies of the cylinder to compute entanglement entropy on non-replicated manifold \cite{Calabrese:2004eu,Cardy:2007mb}
\begin{equation}
  \tr\rho^n_A(t)=
  \frac{\braket{\Psi(x_1,\bar{x}_1)\sigma_n(x_2,\bar{x}_2)\tilde{\sigma}_n(x_3,\bar{x}_3)\Psi^{\dagger}(x_4,\bar{x}_4)}_{C_n}}
  {\left(\braket{\psi(x_1,\bar{x}_1)\psi^{\dagger}(x_4,\bar{x}_4)}_{C_1}\right)^n},
\end{equation}
where $\Psi = \psi_1 \cdot \psi_2 \cdots \psi_n$ is the product of local perturbation operators in each copy of the cylinder. Let $h_\psi$ be the conformal dimension of original perturbation $\psi$, so the conformal dimension of $\Psi$ is $h_\Psi = n h_\psi$. Operator insertion points are 
\begin{equation}
\begin{aligned}
 & x_1=-i\epsilon,\quad x_2=y-t_\omega-t_-,\quad x_3=y+L-t_\omega-t_-,\quad x_4=\hphantom{-}i\epsilon, \\
 & \bar{x}_1=\hphantom{-}i\epsilon,\quad \bar{x}_2=y+t_\omega+t_-,\quad \bar{x}_3=y + L+t_\omega+t_-,\quad \bar{x}_4=-i\epsilon,
\end{aligned}
\end{equation}
and the conformal dimension of twist operators is \cite{Calabrese:2004eu, Cardy:2007mb, PAM_doi:10.1007/JHEP03(2015)163}
\begin{equation}
 \label{eq:conf_dim_twist}
 \Delta_\sigma = 2 H_\sigma = \frac{c}{24} \left(n - \frac{1}{n} \right).
\end{equation}

\subsection{Entanglement Entropy}

\subsubsection{Left boundary}

We will calculate correlators using the same method as in \cite{Caputa:2015waa}. However, instead of single exponential map $\exp(2 \pi x / \beta)$ we will use different maps for holomorphic and antiholomorphic components
\begin{equation}
\begin{aligned}
 \label{eq:exponential2}
 w(x) &= e^{\frac{2 \pi}{\beta_+} x}, \\
 \bar{w}(\bar{x}) &= e^{\frac{2 \pi}{\beta_-} \bar{x}}.
\end{aligned}
\end{equation}
Then the two point function is given by
\begin{equation}
 \braket{\psi(x_1,\bar{x}_1)\psi(x_4,\bar{x}_4)}_{C_1}=\left|\frac{\beta_+ \beta_-}{\pi^2}\sinh\left(\frac{\pi x_{14}}{\beta_+}\right)\sinh\left(\frac{\pi \bar{x}_{14}}{\beta_-}\right)\right|^{-2h_\psi}.
\end{equation}

To calculate 4-point function we combine exponential maps with a M\"obius map
\begin{equation}
 \label{eq:cross-ratio}
 z(w) = \frac{(w_1 - w)w_{34}}{w_{13} (w - w_4)},
\end{equation}
to send insertion points to the standard set of points $w_1 \mapsto 0, w_2 \mapsto z, w_3 \mapsto 1, w_4 \mapsto \infty$ and we define $w_{ij} = w_i - w_j$ and cross ratios are as usual
\begin{equation}
 z = \frac{w_{12} w_{34}}{w_{13} w_{24}}, \qquad 1 - z = \frac{w_{14} w_{23}}{w_{13} w_{24}}.
\end{equation}
This allows us to express the trace of the density matrix in terms of the canonical 4-pt function $G(z, \bar{z})$
\begin{equation}
 \label{eq:oneb}
 \tr \rho^n_A(t) = \left|\frac{\beta_+ \beta_-}{\pi^2 \epsilon_{\mathrm{UV}}^2}\sinh\left(\frac{\pi x_{23}}{\beta_+} \right) \sinh\left(\frac{\pi \bar{x}_{23}}{\beta_-} \right)\right|^{-2H_\sigma}\left|1-z\right|^{4H_\sigma}G(z,\bar{z}),
\end{equation}
where
\begin{align}
 G(z, \bar{z}) &\equiv \lim_{z_4\to \infty}|z_4|^{4h_\Psi} \braket{\psi(z_4,\bar{z}_4)\sigma_n(z,\bar{z})\tilde{\sigma}_n(1,1)\psi(0,0)} \notag \\
&\equiv \braket{\psi | \sigma_n(z,\bar{z})\tilde{\sigma}_n(1,1) | \psi }
\end{align}

The R\'enyi entropies are then given by
\begin{equation}
\label{eq:Renyi_entropy}
S^{(n)}_A=\frac{c(n+1)}{12}\log\left[\frac{\beta_+ \beta_-}{\pi^2 \epsilon_{\mathrm{UV}}^2}\sinh\left(\frac{\pi L}{\beta_+} \right) \sinh\left(\frac{\pi L}{\beta_-} \right) \right] + \frac{1}{n-1}\log\Big[|1-z|^{4H_\sigma}G(z,\bar{z})\Big].
\end{equation}
In the $n \to 1$ limit the first term agrees with the well known formula for the entanglement entropy of rotating BTZ black hole \cite{Hubeny:2007xt} which we will call $S_\text{thermal}$. We denote
\begin{equation}
 S_A = S_\text{thermal} + \Delta S_A.
\end{equation}

The calculation of $G(z, \bar{z})$ proceeds as in the non-rotating BTZ case but with different cross ratios $z$ and $\bar{z}$. We consider CFT in large $c$ limit. Then in the $n \to 1$ limit conformal dimension of twist operators \eqref{eq:conf_dim_twist} goes to $0$. In this regime $G(z, \bar{z})$ does not depend on the details of the CFT and is universal. Such correlators with two heavy and two light operators were studied in \cite{Fitzpatrick:2014vua,Fitzpatrick:2015zha, Asplund:2014coa}. They obtained that
\begin{equation}
  \label{eq:Fitzpatrick}
  \log G(z,\bar{z})\simeq -\frac{c(n-1)}{6}\log\left(\frac{z^{\frac{1}{2}(1-\alpha_\psi)}\bar{z}^{\frac{1}{2}(1-\bar{\alpha}_\psi)}(1-z^{\alpha_\psi})(1-\bar{z}^{\bar{\alpha}_\psi})}{\alpha_\psi\bar{\alpha}_\psi}\right)+{\cal O}((n-1)^2).
\end{equation}
where
\begin{equation}
 \alpha_\psi=\sqrt{1-\frac{24h_\psi}{c}}.
\end{equation}
Note that all information about perturbation is in $\alpha_\psi$ and $z$ does not depend on $h_\psi$.

In the small $\epsilon$ approximation cross-rations are given by
\begin{align}
  z&\simeq 1+\frac{2\pi i \epsilon}{\beta_+}\frac{\sinh\frac{\pi\,L}{\beta_+}}{\sinh\frac{\pi (y+L-t_- -t_\omega)}{\beta_+}\sinh\frac{\pi (y-t_- -t_\omega)}{\beta_+}}+{\cal O}(\epsilon^2), \\
  \bar{z}& \simeq 1-\frac{2\pi i \epsilon}{\beta_-}\frac{\sinh\frac{\pi\,L}{\beta_-}}{\sinh\frac{\pi (y+L+t_- +t_\omega)}{\beta_-}\sinh\frac{\pi (y+t_- +t_\omega)}{\beta_-}}+{\cal O} (\epsilon^2).
\end{align}
So cross-ratios depend on $\beta_+$ and $\beta_-$ respectively but otherwise nothing else has changed compared to $\beta_- = \beta_+$ case.

Since four point function above is approximated with a multi valued function, so we have to be careful when we use these cross-rations in equation \eqref{eq:Renyi_entropy} and pick the correct branch of the logarithm \cite{Asplund:2014coa,Roberts:2014ifa} (also see \cite{Caputa:2015waa}). It is convenient to split the problem into three cases.
\begin{itemize}
 \item Early time: $0 < t_- + t_\omega < y$.
 \item Intermediate time: $y < t_- + t_\omega < y + L$.
 \item Late time: $y + L < t_- + t_\omega$.
\end{itemize}
For antiholomorphic coordinates in all cases we have that $\bar{z}$ are close to 1 and do not contribute to the $\Delta S_A$. On the other hand, imaginary part of $z$ changes the sign in the intermediate time region. In order to obtain non-negative entanglement entropy and be consistent with causality we pick $z \to e^{2 \pi i}$ in the intermediate time and $z \to 1$ in the early and late times.

Therefore, entanglement entropy is given by
\begin{equation}
\Delta S_A = \begin{cases}0\,, \quad \text{ when } t_-+t_\omega < y \,\,\text{or}\,\, t_-+t_\omega > y + L, \\
\dfrac{c}{6}\log\left[\dfrac{\beta_+}{\pi\epsilon}\dfrac{\sin \pi \alpha_\psi}{\alpha_\psi}\dfrac{\sinh\dfrac{\pi(y+L -t_- -t_\omega)}{\beta_+}\sinh\dfrac{\pi(t_- +t_\omega -y)}{\beta_+}}{\sinh \dfrac{\pi L}{\beta_+} }\right] \quad \text{otherwise}\,.
\end{cases}
\label{eq:S_A_CFT2}
\end{equation}

This result can be interpreted as two localized lumps of energy moving in the opposite directions. The lump describing holomorphic mode moves from insertion point at $x = 0$ to $x=y$ at early time. Then at intermediate time it passes region $A$, so we have a non-trivial contribution to entanglement entropy and then at late time the lump of energy leaves the region $A$, so $\Delta S_A = 0$. On the other hand, antiholomorphic mode moves in the opposite direction and never passes through region $A$. So it is expected that $\Delta S_A$ does not depend on $\beta_-$. Note that if we consider negative $y$ and $L$ then only antiholomorphic mode contributes, therefore $\Delta S_A$ does not depend on $\beta_+$.

\subsubsection{Right boundary}

\begin{figure}[htb]
  \centering
  \import{images/}{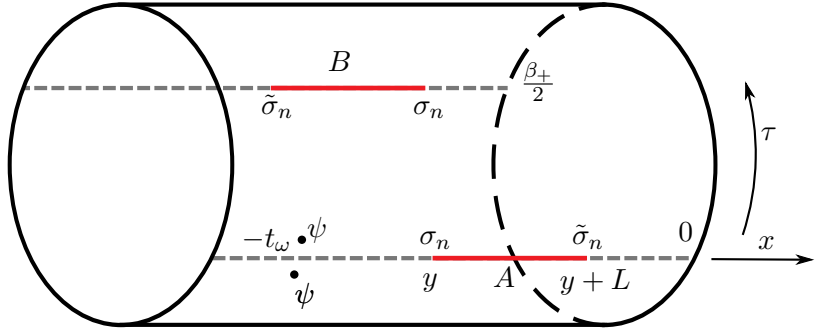}
  \caption{Entanglement region $B$ on the right boundary and operator insertion points on the left boundary depicted on the thermal cylinder.}
  \label{fig:cylinder2}
\end{figure}
% Fix beta in the picture. \beta_\pm?

Calculation for region $B$ on the right boundary is similar. This time twist operators are inserted in the $\text{CFT}_R$ (see \autoref{fig:cylinder2}) so we must add $i \beta_+/2$ or $i \beta_-/2$ to the holomorphic or antiholomorphic coordinates respectively. So the insertion points are
\begin{equation}
\begin{aligned}
 & x_1 = -i\epsilon,\quad x_5 = y+L+\frac{i\beta_+}{2}- t_+-t_\omega,\quad x_6 = y +\frac{i\beta_+}{2}- t_+-t_\omega,\quad x_4 = +i\epsilon \\
 & \bar{x}_1 = +i\epsilon,\quad \bar{x}_5=y+L-\frac{i\beta_-}{2}+ t_++t_\omega,\quad \bar{x}_6=y-\frac{i\beta_-}{2}+ t_++t_\omega,\quad \bar{x}_4 = -i\epsilon\,.
\end{aligned}
\end{equation}

Cross ratios are
\begin{align}
z=z_5&\simeq1-\frac{2\pi i \epsilon}{\beta_+}\frac{\sinh\frac{\pi\,L}{\beta_+}}{\cosh\frac{\pi (y- t_+-t_\omega)}{\beta_+}\cosh\frac{\pi (y+L- t_+-t_\omega)}{\beta_+}}+{\cal O}(\epsilon^2)\,, \\
\bar{z}=\bar{z}_5&\simeq 1+\frac{2\pi i \epsilon}{\beta_-}\frac{\sinh\frac{\pi\,L}{\beta_-}}{\cosh\frac{\pi (y+t_++t_\omega)}{\beta_-}\cosh\frac{\pi (y+L+ t_++t_\omega)}{\beta_-}}+{\cal O}(\epsilon^2)\,.
\end{align}
Note that this time imaginary part does not flip sign for any time. Hence $(z, \bar{z}) \to (1, 1) \quad \forall t_+$. In this case entanglement entropy of region $B$ stays thermal to leading order in $\epsilon$ and
\begin{equation}
 S_B \simeq \frac{c}{6}\log\left[\frac{\beta_+ \beta_-}{\pi^2 \epsilon_{\mathrm{UV}}^2}\sinh\left(\frac{\pi L}{\beta_+} \right) \sinh\left(\frac{\pi L}{\beta_-} \right) \right].
 \label{eq:S_B_CFT}
\end{equation}
Again, this is something that we expect from physical considerations. Perturbation is inserted in the $\mathrm{CFT}_L$. However, the region of entanglement is in the $\mathrm{CFT}_R$. Since both $\mathrm{CFTs}$ do not interact with each other, we expect that entanglement entropy is unchanged at this order in $\epsilon$.

\subsubsection{\texorpdfstring{$S_{A\cup B}$}{S A union B}}

Finally, to find mutual information, we need to calculate $S_{A\cup B}$ and it involves calculating the 6-point function
\begin{equation}
   \tr \rho^n_{A\cup B}=\frac{\braket{\psi(x_1,\bar{x}_1)\sigma_n(x_2,\bar{x}_2)\tilde{\sigma}_n(x_3,\bar{x}_3)\sigma_n(x_5,\bar{x}_5)\tilde{\sigma}_n(x_6,\bar{x}_6)\psi^{\dagger}(x_4,\bar{x}_4)}}{\left(\braket{\psi(x_1,\bar{x}_1)\psi^{\dagger}(x_4,\bar{x}_4)}_{C_1}\right)^n}.
\end{equation}
with operator insertion points
\begin{eqnarray}
x_1&=&-i\epsilon,\quad x_2=y-t_--t_\omega,\quad x_3=y+L-t_--t_\omega,\quad x_4=+i\epsilon \notag \\
\bar{x}_1&=&+i\epsilon,\quad \bar{x}_2=y+t_-+t_\omega,\quad \bar{x}_3=y+L+t_-+t_\omega,\quad \bar{x}_4=-i\epsilon \notag \\
x_5&=&y+L+\frac{i\beta_+}{2}- t_+-t_\omega,\quad x_6=y+\frac{i\beta_+}{2}- t_+-t_\omega,\notag \\
\bar{x}_5&=&y+L-\frac{i\beta_-}{2}+ t_++t_\omega,\quad \bar{x}_6=y-\frac{i\beta_-}{2}+ t_++t_\omega\,.
\end{eqnarray}

After mapping to the plane with exponential map \eqref{eq:exponential2} and another transformation to map insertion points to the four standard points
\begin{equation}
 z(w) = \frac{(w_1 - w)(w_3 - w_4)}{(w_1 - w_3)(w - w_4)},
\end{equation}
we obtain the following 6-point function
\begin{equation}
 \braket{\psi|\sigma_n(z,\bar{z})\tilde{\sigma}_n(1,1)\sigma_n(z_5,\bar{z}_5)\tilde{\sigma}_n(z_6,\bar{z}_6)|\psi}.
\end{equation}

There are two ways in which we can expand 6-point function in terms of 4-point functions.
 $S$-channel
 \begin{align}
 \label{eq:S-channel}
 &\braket{\psi|\sigma_n(z,\bar{z})\tilde{\sigma}_n(1,1)\sigma_n(z_5,\bar{z}_5)\tilde{\sigma}_n(z_6,\bar{z}_6)|\psi} \notag \\
 &=\sum_{\alpha}\braket{\psi|\sigma_n(z,\bar{z})\tilde{\sigma}_n(1,1)|\alpha}\braket{\alpha|\sigma_n(z_5,\bar{z}_5)\tilde{\sigma}_n(z_6,\bar{z}_6)|\psi}
 \end{align}
and $T$-channel
 \begin{align}
 \label{eq:T-channel}
 &\braket{\psi|\sigma_n(z,\bar{z})\tilde{\sigma}_n(1,1)\sigma_n(z_5,\bar{z}_5)\tilde{\sigma}_n(z_6,\bar{z}_6)|\psi} \notag \\
 &=\sum_{\alpha}\braket{\psi|\sigma_n(z,\bar{z})\tilde{\sigma}_n(z_6,\bar{z}_6)|\alpha}\braket{\alpha|\sigma_n(z_5,\bar{z}_5)\tilde{\sigma}_n(1, 1)|\psi}.
 \end{align}
These channels in the CFT correspond to the two ways geodesic can wrap around the black hole horizon \cite{Asplund:2014coa}. So to compute entanglement entropy we need to minimize over both channels.  

For S-channel, we can write an operator product expansion (OPE) of twist operators
\begin{equation}
  \sigma_n(z,\bar{z})\tilde{\sigma}_n(1,1) \sim \mathbb{I} + {\mathcal{O}}\left((z-1)^r\right) \quad \quad r\in \mathbb{N}.
\end{equation}
The dominant contribution in the OPE expansion comes from identity operator. We will ignore all higher order terms. Then summation over $\ket{\alpha}$ reduces to state $\ket{\psi}$ due to orthogonality of 2-point functions
\begin{equation}
  \sum_{\alpha}\braket{\psi|\sigma_n(z,\bar{z})\tilde{\sigma}_n(1,1)|\alpha}\bra{\alpha} \simeq \braket{\psi|\sigma_n(z,\bar{z})\tilde{\sigma}_n(1,1)|\psi}\bra{\psi}.
\end{equation}
The first 4-pt function in equation \eqref{eq:S-channel} is the same as in $S_A$ while the second is the same as in $S_B$. Hence, mutual information $I_{A:B} = S_A + S_B - S_{A \cup B}$ vanishes in $S$-channel.

For $T$-channel, the state $\ket{\psi}$ again yields the dominant term of the eigen-function expansion of 6-point function. Thus, equation \eqref{eq:T-channel} reduces to the product of two 4-point functions
\begin{align}
 \braket{\psi|\sigma_n(z_5,\bar{z}_5)\tilde{\sigma}_n(1,1)|\psi}&=G(z_5,\bar{z}_5), \\
 \braket{\psi|\sigma_n(z,\bar{z})\tilde{\sigma}_n(z_6,\bar{z}_6)|\psi}&=|1-\tilde{z}_2|^{4H_\sigma}|z_2 - z_6|^{-4H_\sigma}G(\tilde{z}_2,\bar{\tilde{z}}_2),
\end{align}
but the cross-ratios are not the same as in $S$-channel:
\begin{align}
z_5&=1-\frac{2\pi i \epsilon}{\beta_+}\frac{\cosh\frac{\pi (t_--t_+)}{\beta_+}}{\sinh\frac{\pi(y+L-t_--t_\omega)}{\beta_+}\cosh\frac{\pi(y+L-t_+-t_\omega)}{\beta_+}}+{\cal O}(\epsilon^2), \\
\bar{z}_5&=1+\frac{2\pi i \epsilon}{\beta_-}\frac{\cosh\frac{\pi (t_--t_+)}{\beta_-}}{\sinh\frac{\pi(y+L+t_-+t_\omega)}{\beta_-}\cosh\frac{\pi(y+L+t_++t_\omega)}{\beta_-}}+{\cal O}(\epsilon^2), \\
\tilde{z}_2&=1+\frac{2\pi i \epsilon}{\beta_+}\frac{\cosh\frac{\pi (t_--t_+)}{\beta_+}}{\sinh\frac{\pi(y-t_--t_\omega)}{\beta_+}\cosh\frac{\pi(y-t_+-t_\omega)}{\beta_+}}+{\cal O}(\epsilon^2), \\
\tilde{\bar{z}}_2&=1-\frac{2\pi i \epsilon}{\beta_-}\frac{\cosh\frac{\pi (t_--t_+)}{\beta_-}}{\sinh\frac{\pi(y+t_-+t_\omega)}{\beta_-}\cosh\frac{\pi(y+t_++t_\omega)}{\beta_-}}+{\cal O}(\epsilon^2).
\end{align}

As before, we need to check monodromies when we use cross-ratios in equation \eqref{eq:Fitzpatrick}. All antiholomorphic cross rations tend to $1$ which we expect because antiholomorphic modes move to the opposite direction and they never reach entanglement region. On the other hand, holomorphic cross ratios are $z_5 \simeq e^{-2 \pi i}$ for $t_- + t_\omega > y+L$ and $\tilde{z}_2 \simeq e^{2\pi i}$ for $t_- + t_\omega > y$.

Just like for $S_A$ and $S_B$, in this case $S_{A \cup B}$ has a thermal piece that comes from mapping cylinder to the plane and the contribution coming from perturbation. The thermal part depends on both $\beta_+$ and $\beta_-$ but the effect of perturbation depends only on $\beta_+$. Let us also introduce notation $\Delta t = t_- - t_+$.

\begingroup
\allowdisplaybreaks
\begin{align}
   S_{A\cup B} &\simeq \frac{c}{3}\log\left|\frac{\beta_+ \beta_-}{\pi^2 \epsilon_{\mathrm{UV}}^2}\cosh\left(\frac{\pi \Delta t}{\beta_+} \right) \cosh\left(\frac{\pi \Delta t}{\beta_-} \right) \right| \qquad \text{when } t_- + t_\omega < y, \\
   S_{A\cup B} &\simeq \frac{c}{3}\log\left|\frac{\beta_+ \beta_-}{\pi^2 \epsilon_{\mathrm{UV}}^2}\cosh\left(\frac{\pi \Delta t}{\beta_+} \right) \cosh\left(\frac{\pi \Delta t}{\beta_-} \right) \right| \notag \\
   &+\frac{c}{6}\log\left(\frac{\beta_+}{\pi\epsilon}\frac{\sin\pi \alpha_\psi}{\alpha_\psi}\frac{\sinh\frac{\pi(t_-+t_w-y)}{\beta_+}\cosh\frac{\pi(t_++t_w-y)}{\beta_+}}{\cosh\frac{\pi \Delta t}{\beta_+}}\right) \quad \text{when }  y<t_-+t_\omega<y+L, \label{eq:saubmid} \\
  S_{A\cup B} &\simeq \frac{c}{3}\log\left|\frac{\beta_+ \beta_-}{\pi^2 \epsilon_{\mathrm{UV}}^2}\cosh\left(\frac{\pi \Delta t}{\beta_+} \right) \cosh\left(\frac{\pi \Delta t}{\beta_-} \right) \right| \notag \\
  & +\frac{c}{3}\log\left(\frac{\beta_+}{\pi\epsilon}\frac{\sin\pi \alpha_\psi}{\alpha_\psi}\right) \notag \\
  & + \frac{c}{6}\log\Bigg(\frac{\sinh\frac{\pi(t_-+t_\omega-y)}{\beta_+}\cosh\frac{\pi(t_++t_\omega-y)}{\beta_+}}{\cosh\frac{\pi \Delta t}{\beta_+}} \Bigg) \notag \\
  & + \frac{c}{6}\log\Bigg(\frac{\sinh\frac{\pi(t_-+t_\omega-y-L)}{\beta_+}\cosh\frac{\pi(t_++t_\omega-y-L)}{\beta_+}}{\cosh\frac{\pi \Delta t}{\beta_+}}\Bigg) \quad \text{when } t_-+t_\omega >y+L, \label{eq:saub}
\end{align}
\endgroup

\subsection{Mutual information and scrambling time}

We can find mutual information $I_{A:B} = S_A + S_B - S_{A \cup B}$.
At early times $t_- + t_\omega$ it is given by

\begin{equation}
 I_{A:B} = \frac{c}{3} \log \frac{\sinh \frac{\pi L}{\beta_+} \sinh \frac{\pi L}{\beta_-}}{\cosh \frac{\pi \Delta t}{\beta_+} \cosh \frac{\pi \Delta t}{\beta_+}}.
\end{equation}
In the region $(y, y+L)$ mutual information evolves as
\begin{equation}
I_{A:B}\simeq I^0_{A:B}+ \frac{c}{6}\log\left[\frac{\sinh\frac{\pi(y+L -t_- -t_\omega)}{\beta_+}\cosh\frac{\pi \Delta t}{\beta_+}}{\cosh\frac{\pi(t_++t_w-y)}{\beta_+}\sinh\frac{\pi\,L}{\beta_+}}\right].
\label{eq:region2}
\end{equation}
It is notable that equation \eqref{eq:region2} is independent of the conformal dimension $h_\psi$.

In the late time regime $t_- + t_\omega > y+L > y$ the mutual information equals
\begin{align}
I_{A:B} &\simeq I^0_{A:B} -\frac{c}{3}\log\left(\frac{\beta_+}{\pi\epsilon}\frac{\sin\pi \alpha_\psi}{\alpha_\psi}\right) \notag \\
&-\frac{c}{6}\log\left(\frac{\sinh\frac{\pi(t_-+t_\omega-y)}{\beta_+}\cosh\frac{\pi(t_++t_\omega-y)}{\beta_+}}{\cosh\frac{\pi \Delta t}{\beta_+}}\frac{\sinh\frac{\pi(t_-+t_\omega-y-L)}{\beta_+}\cosh\frac{\pi(t_++t_\omega-y-L)}{\beta_+}}{\cosh\frac{\pi \Delta t}{\beta_+}}\right).
\label{eq:mutual_information_CFT}
\end{align}

Scrambling time was also investigated \cite{Shenker:2013pqa} where shock-wave geometry in the BTZ background was considered. The authors found mutual information and the time scale when it vanishes which they define to be the scrambling time.

To compare our model with theirs we inserted our perturbation at time $-t_\omega$ in the past and we will evaluate mutual information at $t_- = t_+ = 0$ ($\Delta t = 0$). Then we can find time dependence of mutual information $I_{A:B}(t_\omega)$ and the scrambling time $t_\omega^\star$ is defined by
\begin{equation}
 \label{eq:scrambling_time_equation}
 I_{A:B}(t_\omega^\star) = 0.
\end{equation}.

When calculating scrambling time we are interested in long term behaviour of the system, so we will focus on the case $t_- + t_\omega > y+L > y$. Then equation \eqref{eq:scrambling_time_equation} has the following exact\footnote{It is exact only at this step. We have made some approximations before such as small $\epsilon$.} solution
\begin{equation}
  T = A^2 \cosh \frac{2 \pi L}{\beta_+} + \frac{1}{2} \sinh^2 \frac{2 \pi L}{\beta_+} + \sqrt{ \left( A^2 \cosh \frac{2 \pi L}{\beta_+} + \frac{1}{2} \sinh^2 \frac{2 \pi L}{\beta_+} \right)^2 - A^4 },
\end{equation}
where
\begin{align}
  A &= \frac{2 \pi \alpha_\psi \epsilon \sinh \frac{\pi L}{\beta_+} \sinh \frac{\pi L}{\beta_-} }{\beta_+ \sin(\pi\alpha_\psi)}, \\
  T &= \sinh^2 \frac{2 \pi (t_w^\star - y)}{\beta_+}.
\end{align}

In the limit $t_\omega^\star \gg \beta_+$ we can approximate hyperbolic functions with exponentials. Then scrambling time is given by
\begin{equation}
\label{eq:SCRtime2}
t^\star_\omega=y + \frac{L}{2}-\frac{\beta_+}{2\pi}\log\left(\frac{\beta_+}{\pi\epsilon}\frac{\sin\pi \alpha_\psi}{\alpha_\psi}\right)+\frac{\beta_+}{2\pi}\log\left(4 \sinh\frac{\pi L}{\beta_+}\sinh\frac{\pi L}{\beta_-}\right).
\end{equation}

We can see from \eqref{eq:SCRtime2} that given two black holes with the same Hawking temperature, scrambling time in the rotating BTZ is longer than in the static BTZ  where the dominant term in scrambling time scales with $\beta$ instead of $\beta_+$. Yet, we should recall our previous comment that in the CFT this is caused by the relative placement of entanglement region and the point where perturbation is inserted.

\section{Bulk results}
\label{s:bulk}

\subsection{Free falling particle in the rotating BTZ}

Having calculated the scrambling time in the CFT, we can now calculate it in its holographic dual, rotating BTZ black hole described by the following metric
\begin{equation}
\label{eq:BTZ_metric}
 ds^2 = - \frac{(r^2 - r_+^2)(r^2 - r_-^2)}{R^2 r^2} dt^2 + \frac{R^2 r^2}{(r^2 - r_+^2)(r^2 - r_-^2)} dr^2 + r^2 \left( d\phi - \frac{r_+ r_-}{R r^2}dt \right)^2,
\end{equation}
where $R$ is the radius of $\mathrm{AdS}_3$, $r_\pm$ are radii the outer and inner event horizons.

In the semiclassical approximation we can model the effect of local quench by adding a massive particle to the bulk spacetime. In general, back-reaction of the particle with mass $m$ on the metric is difficult to compute because it requires solving full Einstein's equations. However, in three dimensions general relativity has no local degrees of freedom, so it is possible to compute back-reaction by mapping \cite{hep-th/9901012v2} metric of eternal BTZ to global $\mathrm{AdS}_3$ coordinates $(\rho, \tau, \varphi)$
\begin{equation}
 ds^2 = -(\rho^2 + R^2) d\tau^2 + \frac{R^2}{\rho^2 + R^2} dr^2 + \rho^2 d\varphi^2.
\end{equation}
Back-reaction of particle located at the origin of AdS ($\rho=0$) is known
\begin{equation}
 \label{eq:backreacted_ads}
 ds^2 = -(\rho^2 + R^2 - \mu) d\tau^2 + \frac{R^2}{\rho^2 + R^2 -\mu} d\rho^2 + \rho^2 d\varphi^2.
\end{equation}
where $\mu = 8 G R^2 m$ is just a rescaled mass of a particle. Depending on the value of $\mu$, this geometry describes conical defect or BTZ black hole.

According to Ryu-Takayanagi proposal \cite{Ryu:2006ef, Ryu:2006bv}, holographic entanglement entropy is proportional to the length of geodesic between two points on the boundary of the BTZ spacetime. The length between two boundary points in \eqref{eq:backreacted_ads} was computed in \cite{Nozaki:2013wia} and gives rise to the following entanglement entropy\footnote{In order to compare CFT and bulk results easier, we will write holographic entanglement entropy in terms of boundary central charge $c$ which is related to radius of $\mathrm{AdS}_3$ via $\frac{c}{6} = \frac{R}{4G_N}$.}
\begin{equation}
S_A=\frac{c}{6}\log\left[\frac{2r^{(1)}_\infty\cdot r^{(2)}_{\infty}}{R^2}\frac{\cos\left(|\Delta \tau_{\infty}|a\right)-\cos\left(|\Delta \varphi_{\infty}|a\right)}{a^2}\right],\label{eq:EEbr}
\end{equation}
where $a\equiv \sqrt{1-\frac{\mu}{R^2}}=\alpha_\psi$ carries the information about the perturbation, as in the CFT discussion, $\Delta \tau_{\infty}=\tau^{(2)}_{\infty}-\tau^{(1)}_{\infty}$ and $\Delta \varphi_{\infty}=\varphi^{(2)}_{\infty}-\varphi^{(1)}_{\infty}$ satisfies $0<|\Delta \varphi_\infty|<\pi$.

So our goal in this section will be to find the positions of boundary points in global $\mathrm{AdS}$ and use them to calculate entanglement entropy which exactly matches the CFT result as we will see later in this section. Thus mutual information and scrambling time must agree as well.

In order to find the positions of boundary points, we first need to consider initial conditions of a massive particle and use it to find a map between $\mathrm{BTZ}$ coordinates and the back-reacted geometry in global $\mathrm{AdS}_3$ coordinates.

\subsubsection{Initial conditions}

The perturbation in the boundary CFT did not carry any angular momentum, so we should make sure that the free falling particle has no angular momentum either. Conserved quantities in asymptotically AdS space were considered in \cite{1102.4352v2}. The expression for the angular momentum conserved charge is
\begin{equation}
 L \equiv g_{\mu \nu} (\partial_\varphi)^\mu \dot{z}^\nu (\tau) = \frac{dT}{d\tau} \left[ g_{t\varphi} + g_{\varphi \varphi} \frac{d\Phi}{ dt} \right],
\end{equation}
where trajectory of the particle is $z^\mu (\tau) = (T(\tau), R(\tau), \Phi(\tau))$ and $\tau$ is the proper time. So $L = 0$ corresponds to
\begin{equation}
 \label{eq:phidot}
 \frac{d\Phi}{dt} = - \frac{g_{t \varphi}}{g_{\varphi \varphi}} = \frac{r_- r_+}{R r^2}.
\end{equation}

Similarly, the energy of the particle is
\begin{equation}
 E \equiv -g_{\mu \nu} (\partial_t)^\mu \dot{z}^\nu (\tau) = - \frac{dT}{d\tau} \left[ g_{tt} + g_{t \varphi} \frac{d\Phi}{dt} \right].
\end{equation}
In particular, the energy of zero angular momentum particle in the rotating BTZ background is
\begin{equation}
 E = - \frac{g_{tt}}{\sqrt{ - \left[g_{tt} + g_{\varphi \varphi} \left( \frac{d\Phi}{dt} \right )^2 + g_{t\varphi}\frac{d\Phi}{dt} \right]}} = \sqrt{- g_{tt}}.
\end{equation}
In the second equality the last two terms in the denominator  cancel because of the equation \eqref{eq:phidot}. Hence, the energy of the particle located at $r=\frac{R}{\varepsilon}$ is
\begin{equation}
 E|_{r=R/\varepsilon} = \frac{mR}{\varepsilon} \sqrt{\left(1 - \frac{r_+^2 \varepsilon^2}{R^2}\right) \left(1 - \frac{r_-^2 \varepsilon^2}{R^2}\right)}.
\end{equation}
Note that to leading order the energy is the same as in non-rotating BTZ and matches the energy of local quench in the field theory $E_\mathrm{CFT} = \frac{2h}{\epsilon}$ if massive particle is initially placed at $r = \frac{R}{\varepsilon}$ and its mass is $m = \frac{2h}{R}$.

\subsubsection{Boosts}

BTZ metric \eqref{eq:BTZ_metric} is usually given in terms of dimensionful time $t$. In order to compare bulk results to the CFT we will switch to dimensionless time $t_- = \frac{t}{R}$ (or similarly $t_+$ in the other asymptotic region). Motivated by calculation above and in particular equation \eqref{eq:phidot} we choose initial condition for the particle to be
\begin{align}
   &r(-t_\omega) = \frac{R}{\varepsilon}, \qquad &\phi(-t_\omega) &= 0, \\
   &\dot{r}(-t_\omega) = 0, \quad &\dot{\phi}(-t_\omega) &= \frac{r_- r_+ \varepsilon^2}{R^2}, \quad &\dot{t}_-=1.
\end{align}

In order to find how particle back-reacts on the geometry we first map a moving particle in the rotating BTZ to a static particle in $\mathrm{AdS}_3$. In the non-rotating BTZ case \cite{Caputa:2015waa} we were able to write down the action for a free falling particle and solve the Euler-Lagrange equations explicitly in both Schwarzschild and Kruskal coordinates. The technical reason why it worked was that we could choose $\phi = 0$ at all times. On the other hand, the time evolution of $\phi$ is more complicated in the rotating BTZ due to frame dragging effect. Therefore, following the same strategy and explicitly finding the geodesic in Kruskal coordinates would not work. Instead we will try to work in the $\mathrm{AdS_3}$ embedding coordinates.

The map between embedding coordinates and BTZ coordinates is
\begin{equation}
\begin{aligned}
 \label{eq:embedding2Schwarzschild}
 X_0 &= \pm \sqrt{B(r)} \sinh \tilde{t}, \\
 X_1 &= \hphantom{\pm} \sqrt{A(r)} \cosh \tilde{\phi}, \\
 X_2 &= \hphantom{\pm}\sqrt{A(r)} \sinh \tilde{\phi}, \\
 X_3 &= \pm \sqrt{B(r)} \cosh \tilde{t},
\end{aligned}
\end{equation}
where $+$ and $-$ signs are for left and right asymptotic regions respectively. Functions $A$ and $B$ are
\begin{equation}
\begin{aligned}
 \label{eq:AB_functions}
 A(r) &= R^2 \frac{r^2 - r_-^2}{r_+^2 - r_-^2}, \\
 B(r) &= R^2 \frac{r^2 - r_+^2}{r_+^2 - r_-^2},
\end{aligned}
\end{equation}
and co-rotating coordinates are given by
\begin{align}
 \tilde{\phi} &= \frac{r_+ \phi}{R} + \frac{r_- (t_\pm + t_\omega)}{R}, \\
 \tilde{t} &= \frac{r_+ (t_\pm + t_\omega)}{R} + \frac{r_- \phi}{R}.
\end{align}
Note that we have already shifted time in the co-rotating coordinates by $-t_\omega$ which is equivalent to applying boost in $X_0 - X_3$ plane as was done in \cite{Caputa:2015waa}.

Now we can write initial conditions in the embedding coordinates
\begin{equation}
\begin{aligned}
\label{eq:X1}
  X_0(0) &= 0, & \dot{X}_0(0) &= \sqrt{B(R\varepsilon^{-1})} \, \dot{\tilde{t}}, \\
  X_1(0) &= \sqrt{A(R\varepsilon^{-1})}, & \dot{X}_1(0) &= 0,  \\
  X_2(0) &= 0, & \dot{X}_2(0) &= \sqrt{A(R\varepsilon^{-1})}\,  \dot{\tilde{\phi}}, \\
  X_3(0) &= \sqrt{B(R\varepsilon^{-1})}, & \dot{X}_3(0) &= 0.
\end{aligned}
\end{equation}
Compare it to initial conditions of particle in $\mathrm{AdS}_3$ \footnote{Note different time coordinate $\tau$ in global $\mathrm{AdS_3}$ coordinates, hence different notation for the derivative.}
\begin{equation}
\begin{aligned}
\label{eq:X3}
  X_0(0) &= 0, & X_0^\prime(0) &= R, \\
  X_1(0) &= R, & X_1^\prime(0) &= 0, \\
  X_2(0) &= 0, & X_2^\prime(0) &= 0, \\
  X_3(0) &= 0, & X_3^\prime(0) &= 0.
\end{aligned}
\end{equation}

We can read off the required boosts from equations \eqref{eq:X1}--\eqref{eq:X3}. First, we need a boost in $X_1 - X_3$ plane with
\begin{equation}
 \cosh \lambda_2 = \sqrt{A(R\varepsilon^{-1})} \simeq \frac{R}{\sqrt{r_+^2 - r_-^2}\varepsilon}, \qquad \sinh \lambda_2 = \sqrt{B(R\varepsilon^{-1})} \simeq \frac{R}{\sqrt{r_+^2 - r_-^2}\varepsilon}.
\end{equation}
Hence,
\begin{equation}
 \tanh \lambda_2 = \sqrt{\frac{R^2 - r_+^2 \varepsilon^2}{R^2 - r_-^2 \varepsilon^2}} = \sqrt{\frac{1 - \frac{r_+^2 \varepsilon^2}{R^2}}{1 - \frac{r_-^2 \varepsilon^2}{R^2}}} = \sqrt{1 - \frac{r_+^2 - r_-^2}{R^2} \varepsilon^2} + \mathcal{O}(\varepsilon^2) \approx \sqrt{ 1 - \kappa r_+ \varepsilon^2},
\end{equation}
where $\kappa$ is the surface gravity
\begin{equation}
 \kappa = \frac{r_+^2 - r_-^2}{R^2 r_+}.
\end{equation}

We also need a boost in $X_0-X_2$ plane to account for time derivatives in  \eqref{eq:X1}.
\begin{equation}
 \tanh \lambda_3 = \coth \lambda_2 \frac{r_+ \dot{\phi} + r_- \dot{t}_-}{r_+ \dot{t}_- + r_- \dot{\phi}} = \frac{r_-}{r_+} + \mathcal{O}(\varepsilon).
\end{equation}

It will also be useful to find
\begin{equation}
 \cosh \lambda_3 = \frac{r_+}{\sqrt{r_+^2 - r_-^2}}.
\end{equation}

Note that non-zero but small $\dot{\phi}$ does not affect boost $\lambda_3$ at this order. However, perturbation in the CFT with a spin would contribute to $\lambda_3$ at leading order.

\subsection{Back-reaction map for rotating BTZ}

We will again apply equation \eqref{eq:EEbr} to find the lengths of geodesics. Therefore, we need to map the endpoints of entanglement interval to global AdS coordinates. For our purposes it is enough to find the leading order terms even though the map that maps moving particle to the origin of $\mathrm{AdS_3}$ can be written explicitly. We will compare Schwarzschild and global coordinates via $\mathbb{R}^{2,2}$ embedding coordinates.
\begin{equation}
\begin{aligned}
 \label{eq:global2embedding}
 \sqrt{R^2 + \rho^2} \sin\tau &= X_0 \cosh \lambda_3 - X_2 \sinh \lambda_2, \\
 \sqrt{R^2 + \rho^2} \cos\tau &= X_1 \cosh \lambda_2 - X_3 \sinh \lambda_2, \\
 \rho \sin \varphi &= X_2 \cosh \lambda_3 - X_0 \sinh \lambda_3, \\
 \rho \cos \varphi &= X_3 \cosh \lambda_2 - X_1 \sinh \lambda_2.
\end{aligned}
\end{equation}
We can substitute embedding coordinates in \eqref{eq:global2embedding} from \eqref{eq:embedding2Schwarzschild}. Note that functions from equation \eqref{eq:AB_functions} satisfy $A(r) \simeq B(r)$ at the boundary, so $\cosh \lambda_2 \simeq \sinh \lambda_2$.
\begin{equation}
 \label{eq:map_phi2}
 \tan \varphi = \frac{\sinh \tilde{\phi} \cosh \lambda_3 \mp \sinh \tilde{t} \sinh \lambda_3}{\pm \cosh \tilde{t} \cosh \lambda_2 - \cosh \tilde{\phi} \sinh\lambda_2} \simeq - \frac{\cosh \lambda_3}{\cosh \lambda_2} \frac{\sinh \tilde{\phi} \mp \tanh \lambda_3 \sinh \tilde{t}}{\cosh \tilde{\phi} \mp \cosh \tilde{t}}.
\end{equation}

Similarly,
\begin{equation}
 \label{eq:map_tau2}
 \tan \tau \simeq \frac{\cosh \lambda_3}{\cosh \lambda_2} \cdot \frac{\pm\sinh \tilde{t} - \tanh \lambda_3 \sinh \tilde{\phi}}{\cosh \tilde{\phi} \mp \cosh \tilde{t}}.
\end{equation}
We can obtain the radial coordinate by squaring last two equations in \eqref{eq:global2embedding}

\begin{equation}
 \rho^2 \simeq A^2(r) \left( \big[ \cosh \lambda_2 (\cosh \tilde{\phi} - \cosh \tilde{t}) \big]^2 + \big[ \cosh \lambda_3 (\sinh \tilde{\phi} - \tanh{\lambda_3} \sinh \tilde{t}) \big]^2\right).
\end{equation}
This sum is dominated\footnote{This breaks down in the extremal limit. We will not consider this case here because more approximations break down, e.g. stringy corrections become important \cite{Maldacena:2016hyu} as well.} by the first term as $\cosh \lambda_2 \gg \cosh \lambda_3$. So
\begin{equation}
 \label{eq:map_r2}
 \rho \simeq A(r) \cosh \lambda_2 \left|  \cosh \tilde{\phi} - \cosh \tilde{t} \right|.
\end{equation}

\subsection{Geodesic lengths}
\subsubsection{Geodesic on the left boundary}

As discussed at the beginning of this section, we need to find the positions of endpoints in back-reacted global $\mathrm{AdS}_3$ coordinates. In particular, we need to determine $\rho^{(1)} \rho^{(2)}$, $\Delta \tau$ and $\Delta \varphi$ using the map \eqref{eq:map_r2}, \eqref{eq:map_tau2} and \eqref{eq:map_phi2} given endpoints $(t_-, r_\infty, L_1) = (t_-, Rz_\infty^{-1}, L_1)$ and $(t_-, r_\infty, L_2)$ in rotating BTZ coordinates. Let us also denote the length of the interval by $L = L_2 - L_1$.

The radial coordinates \eqref{eq:map_r2} satisfy
\begin{equation}
 \rho^{(1)} \rho^{(2)} \simeq R^2 \frac{\beta_- \beta_+}{4 z_\infty^2} \cosh^2 \lambda_2 |D_1 D_2|,
\end{equation}
where
\begin{equation}
 D_i = \cosh \tilde{\phi}_i - \cosh \tilde{t}_i,
\end{equation}
and co-rotating coordinates are as before
\begin{align}
 \tilde{\phi}_i &= \frac{r_+ L_i + r_- (t_- + t_\omega)}{R}, \\
 \tilde{t}_i &= \frac{r_+ (t_- + t_\omega) + r_- L_i}{R}.
\end{align}

Similarly, we can write expressions for $\tau$ and $\varphi$ coordinates
\begin{align}
 \tan \tau^{(i)} &\simeq \hphantom{-} \frac{\cosh \lambda_3}{\cosh \lambda_2} \frac{\sinh \tilde{t}_i - \tanh \lambda_3 \sinh \tilde{\phi}_i}{D_i}, \\
 \tan \varphi^{(i)} &\simeq - \frac{\cosh \lambda_3}{\cosh \lambda_2} \frac{\sinh \tilde{\phi}_i - \tanh \lambda_3 \sinh \tilde{t}_i}{D_i}.
\end{align}

Given $\tan \tau$ and $\tan \varphi$ we can find $\tau$ and $\varphi$. However, their values depend on whether $\tilde{\phi}_i > \tilde{t}_i$. We have three cases to consider: early time $0 < t + t_\omega < L_1$, intermediate time $L_1 < t + t_\omega < L_2$ and late time $L_2 < t + t_\omega$.

\paragraph{Early time.}

In this case boundary points are
\begin{align}
 \tau^{(i)} &\simeq \hphantom{\pi -} \ \frac{\cosh \lambda_3}{\cosh \lambda_2} \frac{\sinh \tilde{t}_i - \tanh \lambda_3 \sinh \tilde{\phi}_i}{D_i}, \\
 \varphi^{(i)} &\simeq \pi - \frac{\cosh \lambda_3}{\cosh \lambda_2} \frac{\sinh \tilde{\phi}_i - \tanh \lambda_3 \sinh \tilde{t}_i}{D_i}.
\end{align}

\begin{align}
 |\Delta \tau| & \simeq \frac{\cosh \lambda_3}{\cosh \lambda_2} \frac{1}{D_1 D_2} \big| D_1 (\sinh \tilde{t}_2 - \tanh \lambda_3 \sinh \tilde{\phi}_2) - D_2 (\sinh \tilde{t}_1 - \tanh \lambda_3 \sinh \tilde{\phi}_1) \big|, \label{eq:delta_tau} \\
 |\Delta \varphi| & \simeq \frac{\cosh \lambda_3}{\cosh \lambda_2} \frac{1}{D_1 D_2} \big| D_1 (\sinh \tilde{\phi}_2 - \tanh \lambda_3 \sinh \tilde{t}_2) - D_2 (\sinh \tilde{\phi}_1 - \tanh \lambda_3 \sinh \tilde{t}_1 ) \big|. \label{eq:delta_phi}
\end{align}

We now use the following identity
\begin{align}
 D_1 D_2(|\Delta \varphi|^2 - | \Delta \tau|^2) &= 2 \frac{\cosh^2 \lambda_3}{\cosh^2 \lambda_2} (1 - \tanh^2 \lambda_3) \left[ \cosh ( \tilde{\phi}_2 - \tilde{\phi}_1) - \cosh ( \tilde{t}_2 - \tilde{t}_1) \right] \\
 &= 4 \frac{\cosh^2 \lambda_3}{\cosh^2 \lambda_2} (1 - \tanh^2 \lambda_3) \sinh \frac{\tilde{\phi}_2 - \tilde{\phi}_1 + \tilde{t}_2 - \tilde{t}_1}{2} \sinh\frac{\tilde{\phi}_2 - \tilde{\phi}_1 - \tilde{t}_2 + \tilde{t}_1}{2},
\end{align}
or alternatively
\begin{equation}
  D_1 D_2 (|\Delta \varphi|^2 - | \Delta \tau|^2) = \frac{4}{\cosh^2 \lambda_2} \sinh \frac{\pi L}{\beta_-} \sinh \frac{\pi L}{\beta_+}.
\end{equation}

The length of the geodesic connecting boundary points is
\begin{align}
 L_\gamma &\simeq \log\left[\frac{2\rho^{(1)}\rho^{(2)}}{R^2}\frac{\cos\left(a|\Delta \tau|\right)-\cos\left(a|\Delta \varphi|\right)}{a^2}\right]\simeq\log\left[\frac{\rho^{(1)}\rho^{(2)}}{R^2}(|\Delta \varphi|^2-|\Delta \tau|^2)\right] \\
 &\simeq \log \left[\frac{\beta_- \beta_+}{\pi^2 z_\infty^2} \sinh \frac{\pi L}{\beta_-} \sinh \frac{\pi L}{\beta_+} \right].
\end{align}
This reproduces thermal entanglement entropy answer
\begin{equation}
 S_A = \frac{c}{6} \log \left[\frac{\beta_- \beta_+}{\pi^2 z_\infty^2} \sinh \frac{\pi L}{\beta_-} \sinh \frac{\pi L}{\beta_+} \right].
\end{equation}

\paragraph{Late time.}

Just as in non-rotating BTZ, the points in this case are different but differences between the points are the same as in early time case
\begin{align}
 \tau^{(i)} &\simeq \pi - \frac{\cosh \lambda_3}{\cosh \lambda_2} \frac{\sinh \tilde{t}_i - \tanh \lambda_3 \sinh \tilde{\phi}_i}{|D_i|}, \\
 \varphi^{(i)} &\simeq \hphantom {\pi -} \ \frac{\cosh \lambda_3}{\cosh \lambda_2} \frac{\sinh \tilde{\phi}_i - \tanh \lambda_3 \sinh \tilde{t}_i}{|D_i|}.
\end{align}
Hence, entanglement entropy is thermal
\begin{equation}
 S_A = S_\text{thermal} = \frac{c}{6} \log \left[\frac{\beta_- \beta_+}{\pi^2 z_\infty^2} \sinh \frac{\pi L}{\beta_-} \sinh \frac{\pi L}{\beta_+} \right].
\end{equation}

\paragraph{Intermediate time.}

In this case, the boundary points are
\begin{align}
 \tau^{(1)} & \simeq \pi - \frac{\cosh \lambda_3}{\cosh \lambda_2} \frac{\sinh \tilde{t}_1 - \tanh \lambda_3 \sinh \tilde{\phi}_1}{|D_1|}, \\
 \tau^{(2)} & \simeq \hphantom {\pi -} \ \frac{\cosh \lambda_3}{\cosh \lambda_2} \frac{\sinh \tilde{t}_2 - \tanh \lambda_3 \sinh \tilde{\phi}_2}{D_2}, \\
 \varphi^{(1)} &\simeq \hphantom {\pi -} \  \frac{\cosh \lambda_3}{\cosh \lambda_2} \frac{\sinh \tilde{\phi}_1 - \tanh \lambda_3 \sinh \tilde{t}_1}{|D_1|}, \\
 \varphi^{(2)} &\simeq \pi - \frac{\cosh \lambda_3}{\cosh \lambda_2} \frac{\sinh \tilde{\phi}_2 - \tanh \lambda_3 \sinh \tilde{t}_2}{D_2}.
\end{align}
So,
\begin{align}
 |\Delta \tau| &\simeq \pi - \frac{1}{|D_1| D_2} \big[|D_1| (\sinh \tilde{t}_2 - \tanh \lambda_3 \sinh \tilde{\phi}_2) + D_2 (\sinh \tilde{t}_1 - \tanh \lambda_3 \sinh \tilde{\phi}_1) \big],
 \\
  |\Delta \varphi| &\simeq \pi - \frac{1}{|D_1| D_2} \big[ |D_1| (\sinh \tilde{\phi}_2 - \tanh \lambda_3 \sinh \tilde{t}_2) + D_2 (\sinh \tilde{\phi}_1 - \tanh \lambda_3 \sinh \tilde{t}_1 ) \big].
\end{align}
We can simplify hyperbolic functions to
\begin{align}
 & |D_1| D_2 (|\Delta \varphi| - |\Delta \tau|) \notag \\
 &= \frac{\cosh \lambda_3}{\cosh \lambda_2} (1 + \tanh \lambda_3) \left[ - \sinh(\tilde{t}_2 - \tilde{t}_1) + \sinh(\tilde{\phi}_2 - \tilde{t}_1) - \sinh(\tilde{\phi}_2 - \tilde{\phi}_1) + \sinh(\tilde{t}_2 - \tilde{\phi}_1) \right] \notag \\
 &= 4 \frac{\cosh \lambda_3}{\cosh \lambda_2} (1 + \tanh \lambda_3) \sinh\frac{\tilde{t}_1 - \tilde{\phi}_1}{2} \sinh\frac{\tilde{t}_2 - \tilde{\phi}_2}{2} \sinh\frac{\tilde{t}_1 + \tilde{\phi}_1 - \tilde{t}_2 - \tilde{\phi}_2}{2}.
\end{align}
Let us substitute the values of co-rotating coordinates
\begin{align}
  & |D_1| D_2 (|\Delta \varphi| - |\Delta \tau|) \notag \\
  &= 4 \frac{\cosh \lambda_3}{\cosh \lambda_2} (1 + \tanh \lambda_3) \sinh \frac{\pi(t_\omega + t_- - L_1)}{\beta_+} \sinh \frac{\pi(t_\omega + t_- - L_2)}{\beta_+} \sinh \frac{\pi(L_1 - L_2)}{\beta_-} \notag \\
  & = 4 \frac{\cosh \lambda_3}{\cosh \lambda_2} (1 + \tanh \lambda_3) \sinh \frac{\pi(t_\omega + t_- - L_1)}{\beta_+} \sinh \frac{\pi(L_2-t_\omega - t_-)}{\beta_+} \sinh \frac{\pi L}{\beta_-}.
\end{align}
We can rewrite this as
\begin{align}
  \delta :&= |\Delta \varphi| - |\Delta \tau| \notag \\
  & = \frac{4}{|D_1| D_2} \frac{\cosh \lambda_3}{\cosh \lambda_2} (1 + \tanh \lambda_3)\sinh \frac{\pi L}{\beta_-} \sinh \frac{\pi L}{\beta_+} \cdot \frac{\sinh \frac{\pi(t_\omega + t_- - L_1)}{\beta_+} \sinh \frac{\pi(L_2-t_\omega - t_-)}{\beta_+}}{\sinh \frac{\pi L}{\beta_
  +}}.
\end{align}
Now we can calculate the length of the geodesic connecting the two points
\begin{align}
L_\gamma &\simeq \log\left[\frac{2\rho^{(1)}\rho^{(2)}}{R^2}\frac{\cos\left(a|\Delta \tau|\right)-\cos\left(a|\Delta \varphi|\right)}{a^2}\right] \\
& \simeq\log\left[\frac{ 2 \rho^{(1)}\rho^{(2)}}{R^2}\frac{\sin\pi a}{a}\delta\right] \\
& \simeq\log\left[\frac{\beta_- \beta_+}{\pi^2 z_\infty^2} \sinh \frac{\pi L}{\beta_-} \sinh \frac{\pi L}{\beta_+}  \right] \notag \\
&+ \log \left[2 \cosh \lambda_2 \cosh \lambda_3 (1 + \tanh \lambda_3) \frac{\sin\pi a}{a}\frac{\sinh\frac{\pi(t_\omega+t_--L_1)}{\beta_+}\sinh\frac{\pi(L_2-t_\omega-t_-)}{\beta_+}}{\sinh\frac{\pi L}{\beta_+}}
\right].
\end{align}
Note that
\begin{equation}
 2 \cosh \lambda_2 \cosh \lambda_3 (1 + \tanh \lambda_3) = 2 \frac{R}{\epsilon \sqrt{r_+^2 - r_-^2}} \frac{r_+}{\sqrt{r_+^2 - r_-^2}} \frac{r_+ + r_-}{r_+} = \frac{\beta_+}{\pi \epsilon}.
\end{equation}
Thus entanglement entropy is
\begin{equation}
 S_A = S_\text{thermal} + \frac{c}{6} \log \left[\dfrac{\beta_+}{\pi \epsilon} \dfrac{\sin\pi a}{a}\frac{\sinh\dfrac{\pi(t_\omega+t_--L_1)}{\beta_+}\sinh\dfrac{\pi(L_2-t_\omega-t_-)}{\beta_+}}{\sinh\dfrac{\pi L}{\beta_+}}
\right].
\end{equation}
So holographic entanglement entropy precisely matches the CFT result \eqref{eq:S_A_CFT2}.

\subsubsection{Geodesic on the right boundary}

The two endpoints of the entanglement region $B$ in the right boundary are $(t_+,r_\infty,L_1)$ and $(t_+,r_\infty,L_2)$. Their radial coordinates satisfy
\begin{equation}
 \rho^{(1)} \rho^{(2)} \simeq R^2 \frac{\beta_- \beta_+}{4 z_\infty^2} \cosh^2 \lambda_2 |D_1 D_2|,
\end{equation}
where
\begin{equation}
D_i = \cosh\tilde{\phi}_i + \cosh\tilde{t}_i \equiv \cosh \left( \frac{r_+ L_i}{R} + \frac{r_- (t_+ + t_\omega)}{R} \right) + \cosh \left( \frac{r_+ (t_+ + t_\omega)}{R} + \frac{r_- L_i}{R}\right).
\end{equation}

The other coordinates satisfy
\begin{align}
 \tan \tau^{(i)}  \simeq - \frac{\cosh \lambda_3}{\cosh \lambda_2} \frac{\sinh \tilde{t}_i + \tanh \lambda_3 \sinh \tilde{\phi}_i}{D_i},  \\
 \tan \varphi^{(i)} \simeq - \frac{\cosh \lambda_3}{\cosh \lambda_2} \frac{\sinh \tilde{\phi}_i + \tanh \lambda_3 \sinh \tilde{t}_i}{D_i}.
\end{align}
On the right boundary $D_i > 0$, so we always have
\begin{align}
 \tau^{(i)}  \simeq \pi - \frac{\cosh \lambda_3}{\cosh \lambda_2} \frac{\sinh \tilde{t}_i + \tanh \lambda_3 \sinh \tilde{\phi}_i}{D_i},  \\
 \varphi^{(i)} \simeq \pi - \frac{\cosh \lambda_3}{\cosh \lambda_2} \frac{\sinh \tilde{\phi}_i + \tanh \lambda_3 \sinh \tilde{t}_i}{D_i}.
\end{align}
Calculation proceeds as in early time on the left boundary with $\lambda_3 \mapsto -\lambda_3$. But early time case did not depend on $\lambda_3$ anyway.

Geodesic length is
\begin{align}
 L_\gamma &\simeq \log\left[\frac{2 \rho^{(1)} \rho^{(2)}}{R^2}\frac{\cos\left(a|\Delta \tau|\right)-\cos\left(a|\Delta \varphi|\right)}{a^2}\right]\simeq\log\left[\frac{\rho^{(1)} \rho^{(2)}}{R^2}(|\Delta \varphi|^2-|\Delta \tau|^2)\right] \\
 & \simeq \log \left[\frac{\beta_- \beta_+}{\pi^2 z_\infty^2} \sinh \frac{\pi L}{\beta_-} \sinh \frac{\pi L}{\beta_+} \right].
\end{align}
This again reproduces thermal entanglement entropy answer
\begin{equation}
 S_B = \frac{c}{6} \log \left[\frac{\beta_- \beta_+}{\pi^2 z_\infty^2} \sinh \frac{\pi L}{\beta_-} \sinh \frac{\pi L}{\beta_+} \right].
\end{equation}

\subsubsection{Geodesics across the horizon}

\begin{figure}[htb]
 \centering
 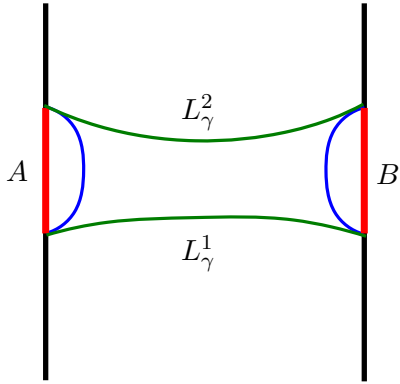
 \caption{Spatial cross-section of BTZ. There is a competition between green geodesics across the horizon and blue geodesics connecting end points on the same boundary.}
 \label{fig:geodesics}
\end{figure}

We will repeat the calculation with points on the different boundaries. In this case geodesic connects points $(t_\mp, r_\infty, L_i)$ (the first point is on the left boundary, the second is on the right). Possible geodesics are shown in \autoref{fig:geodesics}. In order to find entanglement entropy, we need to compare the lengths of geodesics across the horizon and geodesics on the same boundary. Then entanglement entropy of $S_{A \cup B}$ will be given by the set of geodesics with the smaller length. This mirrors $S$ and $T$-channel expansion in the CFT and ensures that mutual information is non-negative.

The product of radial coordinates satisfies
\begin{equation}
 \rho^{(1)} \rho^{(2)} \simeq R^2 \frac{\beta_- \beta_+}{4 z_\infty^2} \cosh^2 \lambda_2 |D_1 D_2|,
\end{equation}
as before but now $D_1$ and $D_2$ have different expressions. We will add $\pm$ superscripts to keep track of which boundary co-rotating coordinates belong to
\begin{align}
 D_1 = \cosh \tilde{\phi}_i^- - \cosh \tilde{t}_i^-, \\
 D_2 = \cosh \tilde{\phi}^+_i + \cosh \tilde{t}^+_i.
\end{align}

\paragraph{Points on the boundary.}At early times $L_i > t_- + t_\omega$ points on the left boundary satisfy
\begin{align}
 \tau^{(1)} &\simeq \hphantom{\pi -} \ \frac{\cosh \lambda_3}{\cosh \lambda_2} \frac{\sinh \tilde{t}_i^- - \tanh \lambda_3 \sinh \tilde{\phi}_i^-}{D_1}, \\
 \varphi^{(1)} &\simeq \pi - \frac{\cosh \lambda_3}{\cosh \lambda_2} \frac{\sinh \tilde{\phi}_i^- - \tanh \lambda_3 \sinh \tilde{t}_i^-}{D_1}.
\end{align}
whereas at late times
\begin{align}
 \tau^{(1)} & \simeq  \pi - \frac{\cosh \lambda_3}{\cosh \lambda_2} \frac{\sinh \tilde{t}_i^- - \tanh \lambda_3 \sinh \tilde{\phi}_i^-}{D_1}, \\
 \varphi^{(1)} &  \simeq \hphantom{\pi -} \ \frac{\cosh \lambda_3}{\cosh \lambda_2} \frac{\sinh \tilde{\phi}_i^- - \tanh \lambda_3 \sinh \tilde{t}_i^-}{D_1}.
\end{align}
Similarly, coordinates boundary points on the right are always
\begin{align}
 \tau^{(2)} & \simeq \hphantom{\pi} - \frac{\cosh \lambda_3}{\cosh \lambda_2} \frac{\sinh \tilde{t}_i^+ + \tanh \lambda_3 \sinh \tilde{\phi}_i^+}{D_2}, \\
 \varphi^{(2)} &  \simeq \pi - \frac{\cosh \lambda_3}{\cosh \lambda_2} \frac{\sinh \tilde{\phi}_i^+ + \tanh \lambda_3 \sinh \tilde{t}_i^+}{D_2}.
\end{align}

\paragraph{Early time.} At early times $t_- + t_\omega < L_i$ we have

\begin{equation}
\begin{aligned}
 |\Delta \tau| &= |\tau^{(1)}-\tau^{(2)}| \simeq \frac{1}{D_1 D_2} \frac{\cosh \lambda_3}{\cosh \lambda_2}\left|D_2( \sinh \tilde{t}_i^- - \tanh \lambda_3 \sinh \tilde{\phi}_i^- ) + D_1 (\sinh \tilde{t}_i^+ + \tanh \lambda_3 \sinh \tilde{\phi}_i^+) \right|\,, \\
 |\Delta \varphi| &= |\varphi^{(1)} - \varphi^{(2)}| \simeq \frac{1}{D_1 D_2} \frac{\cosh \lambda_3}{\cosh \lambda_2} \left| D_2 (\sinh \tilde{\phi}_i^- - \tanh \lambda_3 \sinh \tilde{t}_i^-) - D_1 (\sinh \tilde{\phi}_i^+ + \tanh \lambda_3 \sinh \tilde{t}_i^+ ) \right|.
\end{aligned}
\end{equation}

We now use the following identity
\begin{align}
 D_1 D_2(|\Delta \varphi|^2 - | \Delta \tau|^2) &= 2 \frac{\cosh^2 \lambda_3}{\cosh^2 \lambda_2} (1 - \tanh^2 \lambda_3) \left[ \cosh ( \tilde{\phi}_i^- - \tilde{\phi}_i^+) + \cosh ( \tilde{t}_i^- - \tilde{t}_i^+) \right] \\
 &= \frac{4}{\cosh^2 \lambda_2} \cosh \frac{\tilde{\phi}_i^- - \tilde{\phi}_i^+ + \tilde{t}_i^- - \tilde{t}_i^+}{2} \cosh \frac{\tilde{\phi}_i^- - \tilde{\phi}_i^+ - \tilde{t}_i^- + \tilde{t}_i^+}{2}.
\end{align}
Let $\Delta t = t_- - t_+$, then
\begin{equation}
  D_1 D_2 (|\Delta \varphi|^2 - | \Delta \tau|^2) = \frac{4}{\cosh^2 \lambda_2} \sinh \frac{\pi \Delta t}{\beta_-} \sinh \frac{\pi \Delta t}{\beta_+}.
\end{equation}

Plugging this into the geodesic length equation \eqref{eq:EEbr}, we obtain that at early times
\begin{equation}
 S_{A \cup B} = \frac{c}{6} \log \left[\frac{\beta_- \beta_+}{\pi^2 z_\infty^2} \cosh \frac{\pi \Delta t}{\beta_-} \cosh \frac{\pi \Delta t}{\beta_+} \right].
\end{equation}

\paragraph{Late time.} In the late time regime $t_- + t_\omega > L_i$ we have

\begin{equation}
\begin{aligned}
 |\Delta \tau| &= |\tau^{(1)}-\tau^{(2)}| \simeq \pi - \frac{1}{|D_1| D_2} \frac{\cosh \lambda_3}{\cosh \lambda_2}\left|D_2( \sinh \tilde{t}_i^- - \tanh \lambda_3 \sinh \tilde{\phi}_i^- ) - |D_1| (\sinh \tilde{t}_i^+ + \tanh \lambda_3 \sinh \tilde{\phi}_i^+) \right|\,, \\
 |\Delta \varphi| &= |\varphi^{(1)} - \varphi^{(2)}| \simeq \pi - \frac{1}{|D_1| D_2} \frac{\cosh \lambda_3}{\cosh \lambda_2} \left| D_2 (\sinh \tilde{\phi}_i^- - \tanh \lambda_3 \sinh \tilde{t}_i^-) + |D_1| (\sinh \tilde{\phi}_i^+ + \tanh \lambda_3 \sinh \tilde{t}_i^+ ) \right|.
\end{aligned}
\end{equation}

As in the non-rotating case, $|\Delta \tau|$ and $|\Delta \varphi|$ are close to each other, so we define $\delta = |\Delta \varphi| - |\Delta \tau|$. Then
\begin{align}
 |D_1| D_2 &(|\Delta \varphi| - |\Delta \tau) \\
 &= 2 \frac{\cosh \lambda_3}{\cosh \lambda_2} (1 + \tanh \lambda_3) \left[ \sinh (\tilde{t}_i^- - \tilde{t}_i^+) + \sinh (\tilde{t}_i^- - \tilde{\phi}_i^+)  - \sinh(\tilde{\phi}_i^- - \tilde{\phi}_i^+) - \sinh (\tilde{\phi}_i^- - \tilde{t}_i^+)\right] \\
&= 4 \frac{\cosh \lambda_3}{\cosh \lambda_2} (1 + \tanh \lambda_3) \cosh \frac{\tilde{t}^-_i + \tilde{\phi}^-_i - \tilde{t}^+_i - \tilde{\phi}^+_i}{2} \cosh \frac{\tilde{t}^+_i - \tilde{\phi}_i^+}{2} \sinh \frac{\tilde{t}^-_i - \tilde{\phi}_i^-}{2}.
\end{align}

\begin{equation}
 |D_1| D_2 (|\Delta \varphi| - |\Delta \tau) = 4 \frac{\cosh \lambda_3}{\cosh \lambda_2} (1 + \tanh \lambda_3) \cosh \frac{\pi \Delta t}{\beta_-} \cosh \frac{\pi (t_+ + t_\omega - L_i)}{\beta_+} \sinh \frac{\pi (t_- + t_\omega - L_i)}{\beta_+}.
\end{equation}

Therefore, the lengths of geodesics are

\begin{align}
L^1_\gamma&\simeq \log\left[\frac{\beta_- \beta_+\cosh\frac{\pi\Delta t}{\beta_-} \cosh\frac{\pi\Delta t}{\beta_+}}{\pi^2 z_\infty^2}
\frac{\beta_+}{\pi\epsilon}\frac{\sin\pi a}{a}\frac{\sinh\frac{\pi(t_-+t_\omega-L_1)}{\beta_+}\cosh\frac{\pi(t_++t_\omega-L_1)}{\beta_+}}{\cosh\frac{\pi\Delta t}{\beta_+}}\right], \\
L^2_\gamma&\simeq \log\left[\frac{\beta_- \beta_+\cosh\frac{\pi\Delta t}{\beta_-} \cosh\frac{\pi\Delta t}{\beta_+}}{\pi^2 z_\infty^2}
\frac{\beta_+}{\pi\epsilon}\frac{\sin\pi a}{a}\frac{\sinh\frac{\pi(t_-+t_\omega-L_2)}{\beta_+}\cosh\frac{\pi(t_++t_\omega-L_2)}{\beta_+}}{\cosh\frac{\pi\Delta t}{\beta_+}}\right].
\end{align}

Finally, the entanglement entropy is given by Ryu-Takayanagi formula
\begin{equation}
S_{A\cup B}\simeq \min \begin{cases}
                        \frac{c}{6}\left(L^1_\gamma+L^2_\gamma\right), \\
                        S_A + S_B.
                  \end{cases}
\end{equation}
which matches with the CFT result.

\paragraph{Intermediate time. } As in \cite{Caputa:2015waa}, the intermediate time case is a combination of early and late times cases. The lengths of each geodesic are,
\begin{align}
L^1_\gamma&\simeq \log\left[\frac{\beta_- \beta_+}{\pi^2 z_\infty^2} \cosh\frac{\pi\Delta t}{\beta_-} \cosh\frac{\pi\Delta t}{\beta_+}
\frac{\beta_+}{\pi\epsilon}\frac{\sin\pi a}{a}\frac{\sinh\frac{\pi(t_-+t_\omega-L_1)}{\beta_+}\cosh\frac{\pi(t_++t_\omega - L_1)}{\beta_+}}{\cosh\frac{\pi\Delta t}{\beta_+}}\right], \\
L^2_\gamma &\simeq \log\left[\frac{\beta_- \beta_+}{\pi^2 z_\infty^2} \cosh\frac{\pi\Delta t}{\beta_-} \cosh\frac{\pi\Delta t}{\beta_+} \right].
\end{align}

All holographic entanglement entropies in the rotating BTZ case exactly match their CFT equivalents, so bulk mutual information and the scrambling time will also match the CFT result \eqref{eq:SCRtime2}.

\section{Summary}
In this paper we discussed a massive free-falling particle in the rotating BTZ background. We calculated particle's back-reaction on the geometry and used it to find entanglement entropy and scrambling time. We compared our results with the CFT calculation and found perfect matching.

One of the possible future directions would be to consider how particle with a spin back-reacts on BTZ geometry.

\paragraph{Acknowledgements.} I am very grateful to my supervisor Joan Simón for useful discussions and guidance during my PhD. I would also like to thank Paweł Caputa, J.~S, Tadashi Takayanagi and Kento Watanabe for collaborating with me and giving the opportunity to learn more about this subject. I was supported by the University of Edinburgh Principal's Career Development PhD scholarship.

\bibliographystyle{utphys}
\bibliography{rotating}

\providecommand{\href}[2]{#2}\begingroup\raggedright\begin{thebibliography}{10}

\bibitem{Maldacena:1997re}
J.~M. Maldacena, ``{The Large N limit of superconformal field theories and
  supergravity},'' \href{http://dx.doi.org/10.1023/A:1026654312961}{{\em Int.
  J. Theor. Phys.} {\bfseries 38} (1999) 1113--1133},
  \href{http://arxiv.org/abs/hep-th/9711200}{{\ttfamily arXiv:hep-th/9711200
  [hep-th]}}. [Adv. Theor. Math. Phys.2,231(1998)].

\bibitem{vonNeumann:1927:TQG}
J.~von Neumann, ``{Thermodynamik quantenmechanischer Gesamtheiten},'' {\em
  {Nachrichten von der Gesellschaft der Wissenschaften zu G{\"o}ttingen}}
  {\bfseries 1} (1927) 273--291.

\bibitem{Ryu:2006bv}
S.~Ryu and T.~Takayanagi, ``{Holographic derivation of entanglement entropy
  from AdS/CFT},'' \href{http://dx.doi.org/10.1103/PhysRevLett.96.181602}{{\em
  Phys.Rev.Lett.} {\bfseries 96} (2006) 181602},
  \href{http://arxiv.org/abs/hep-th/0603001}{{\ttfamily arXiv:hep-th/0603001
  [hep-th]}}.

\bibitem{Ryu:2006ef}
S.~Ryu and T.~Takayanagi, ``{Aspects of Holographic Entanglement Entropy},''
  \href{http://dx.doi.org/10.1088/1126-6708/2006/08/045}{{\em JHEP} {\bfseries
  0608} (2006) 045}, \href{http://arxiv.org/abs/hep-th/0605073}{{\ttfamily
  arXiv:hep-th/0605073 [hep-th]}}.

\bibitem{0808.2096v1}
Y.~Sekino and L.~Susskind, ``{Fast Scramblers},''
  \href{http://dx.doi.org/10.1088/1126-6708/2008/10/065}{{\em JHEP} {\bfseries
  0810} (Aug., 2008) 065}, \href{http://arxiv.org/abs/0808.2096v1}{{\ttfamily
  arXiv:0808.2096v1 [hep-th]}}.

\bibitem{Shenker:2013pqa}
S.~H. Shenker and D.~Stanford, ``{Black holes and the butterfly effect},''
  \href{http://dx.doi.org/10.1007/JHEP03(2014)067}{{\em JHEP} {\bfseries 03}
  (2014) 067}, \href{http://arxiv.org/abs/1306.0622}{{\ttfamily arXiv:1306.0622
  [hep-th]}}.

\bibitem{Shenker:2013yza}
S.~H. Shenker and D.~Stanford, ``{Multiple Shocks},''
  \href{http://dx.doi.org/10.1007/JHEP12(2014)046}{{\em JHEP} {\bfseries 12}
  (2014) 046}, \href{http://arxiv.org/abs/1312.3296}{{\ttfamily arXiv:1312.3296
  [hep-th]}}.

\bibitem{Roberts:2014isa}
D.~A. Roberts, D.~Stanford, and L.~Susskind, ``{Localized shocks},''
  \href{http://dx.doi.org/10.1007/JHEP03(2015)051}{{\em JHEP} {\bfseries 03}
  (2015) 051}, \href{http://arxiv.org/abs/1409.8180}{{\ttfamily arXiv:1409.8180
  [hep-th]}}.

\bibitem{Reynolds:2146285}
A.~P. Reynolds and S.~F. Ross, ``{Butterflies with rotation and charge},''
  \href{http://dx.doi.org/10.1088/0264-9381/33/21/215008}{{\em Classical and
  Quantum Gravity} {\bfseries 33} no.~21, (Apr., 2016) 215008},
  \href{http://arxiv.org/abs/1604.04099v1}{{\ttfamily 1604.04099v1 [hep-th]}}.

\bibitem{Leichenauer:2014nxa}
S.~Leichenauer, ``{Disrupting Entanglement of Black Holes},''
  \href{http://dx.doi.org/10.1103/PhysRevD.90.046009}{{\em Phys. Rev.}
  {\bfseries D90} no.~4, (2014) 046009},
  \href{http://arxiv.org/abs/1405.7365}{{\ttfamily arXiv:1405.7365 [hep-th]}}.

\bibitem{Caputa:2014eta}
P.~Caputa, J.~Sim{\'o}n, A.~{\v S}tikonas, and T.~Takayanagi, ``{Quantum
  Entanglement of Localized Excited States at Finite Temperature},''
  \href{http://dx.doi.org/10.1007/JHEP01(2015)102}{{\em JHEP} {\bfseries 01}
  (2015) 102}, \href{http://arxiv.org/abs/1410.2287}{{\ttfamily arXiv:1410.2287
  [hep-th]}}.

\bibitem{PAM_doi:10.1007/JHEP08(2016)106}
J.~Maldacena, S.~H. Shenker, and D.~Stanford, ``{A bound on chaos},''
  \href{http://dx.doi.org/10.1007/JHEP08(2016)106}{{\em {Journal of High Energy
  Physics}} {\bfseries 2016} no.~8, (2016) 1--17},
  \href{http://arxiv.org/abs/1503.01409}{{\ttfamily arXiv:1503.01409
  [hep-th]}}.

\bibitem{1412.6087v3}
S.~H. Shenker and D.~Stanford, ``{Stringy effects in scrambling},'' Mar., 2015.

\bibitem{Caputa:2015waa}
P.~Caputa, J.~Sim{\'o}n, A.~{\v S}tikonas, T.~Takayanagi, and K.~Watanabe,
  ``{Scrambling time from local perturbations of the eternal BTZ black hole},''
  \href{http://dx.doi.org/10.1007/JHEP08(2015)011}{{\em JHEP} {\bfseries 08}
  (2015) 011}, \href{http://arxiv.org/abs/1503.08161}{{\ttfamily
  arXiv:1503.08161 [hep-th]}}.

\bibitem{Caputa:2015qbk}
P.~Caputa, M.~Nozaki, and T.~Numasawa, ``{Charged Entanglement Entropy of Local
  Operators},'' \href{http://dx.doi.org/10.1103/PhysRevD.93.105032}{{\em Phys.
  Rev.} {\bfseries D93} no.~10, (2016) 105032},
  \href{http://arxiv.org/abs/1512.08132}{{\ttfamily arXiv:1512.08132
  [hep-th]}}.

\bibitem{Caputa:2017ixa}
P.~Caputa, S.~R. Das, M.~Nozaki, and A.~Tomiya, ``{Quantum Quench and Scaling
  of Entanglement Entropy},''
  \href{http://dx.doi.org/10.1016/j.physletb.2017.06.017}{{\em Phys. Lett.}
  {\bfseries B772} (2017) 53--57},
  \href{http://arxiv.org/abs/1702.04359}{{\ttfamily arXiv:1702.04359
  [hep-th]}}.

\bibitem{BenTov:2017kyf}
Y.~BenTov and J.~Swearngin, ``{Gravitational shockwaves on rotating black
  holes},'' \href{http://arxiv.org/abs/1706.03430}{{\ttfamily arXiv:1706.03430
  [gr-qc]}}.

\bibitem{0704.3906v2}
M.~M. Wolf, F.~Verstraete, M.~B. Hastings, and J.~I. Cirac, ``{Area laws in
  quantum systems: mutual information and correlations},''
  \href{http://dx.doi.org/10.1103/PhysRevLett.100.070502}{{\em Phys. Rev.
  Lett.} {\bfseries 100} (Mar., 2008) 070502},
  \href{http://arxiv.org/abs/0704.3906v2}{{\ttfamily arXiv:0704.3906v2
  [quant-ph]}}.

\bibitem{Maldacena:2001kr}
J.~M. Maldacena, ``{Eternal black holes in anti-de Sitter},''
  \href{http://dx.doi.org/10.1088/1126-6708/2003/04/021}{{\em JHEP} {\bfseries
  0304} (2003) 021}, \href{http://arxiv.org/abs/hep-th/0106112}{{\ttfamily
  arXiv:hep-th/0106112 [hep-th]}}.

\bibitem{Hartman:2013qma}
T.~Hartman and J.~Maldacena, ``{Time Evolution of Entanglement Entropy from
  Black Hole Interiors},''
  \href{http://dx.doi.org/10.1007/JHEP05(2013)014}{{\em JHEP} {\bfseries 1305}
  (2013) 014}, \href{http://arxiv.org/abs/1303.1080}{{\ttfamily arXiv:1303.1080
  [hep-th]}}.

\bibitem{PAM_doi:10.1007/JHEP11(2013)052}
P.~Caputa, G.~Mandal, and R.~Sinha, ``{Dynamical entanglement entropy with
  angular momentum and U(1) charge},''
  \href{http://dx.doi.org/10.1007/JHEP11(2013)052}{{\em {Journal of High Energy
  Physics}} {\bfseries 2013} no.~11, (2013) 1--25}.

\bibitem{Hubeny:2007xt}
V.~E. Hubeny, M.~Rangamani, and T.~Takayanagi, ``{A Covariant holographic
  entanglement entropy proposal},''
  \href{http://dx.doi.org/10.1088/1126-6708/2007/07/062}{{\em JHEP} {\bfseries
  0707} (2007) 062}, \href{http://arxiv.org/abs/0705.0016}{{\ttfamily
  arXiv:0705.0016 [hep-th]}}.

\bibitem{PAM_doi:10.1007/JHEP01(2015)036}
G.~Mandal, R.~Sinha, and N.~Sorokhaibam, ``{The inside outs of AdS3/CFT2: exact
  AdS wormholes with entangled CFT duals},''
  \href{http://dx.doi.org/10.1007/JHEP01(2015)036}{{\em {Journal of High Energy
  Physics}} {\bfseries 2015} no.~1, (2015) 1--40}.
  \url{http://dx.doi.org/10.1007/JHEP01(2015)036}.

\bibitem{1211.2887v2}
I.~A. Morrison and M.~M. Roberts, ``{Mutual information between thermo-field
  doubles and disconnected holographic boundaries},''
  \href{http://dx.doi.org/10.1007/JHEP07(2013)081}{{\em JHEP} (June, 2013)
  081}, \href{http://arxiv.org/abs/1211.2887v2}{{\ttfamily arXiv:1211.2887v2
  [hep-th]}}.

\bibitem{Calabrese:2004eu}
P.~Calabrese and J.~L. Cardy, ``{Entanglement entropy and quantum field
  theory},'' \href{http://dx.doi.org/10.1088/1742-5468/2004/06/P06002}{{\em J.
  Stat. Mech.} {\bfseries 0406} (2004) P06002},
  \href{http://arxiv.org/abs/hep-th/0405152}{{\ttfamily arXiv:hep-th/0405152
  [hep-th]}}.

\bibitem{Cardy:2007mb}
J.~L. Cardy, O.~A. Castro-Alvaredo, and B.~Doyon, ``{Form factors of
  branch-point twist fields in quantum integrable models and entanglement
  entropy},'' \href{http://dx.doi.org/10.1007/s10955-007-9422-x}{{\em J.
  Statist. Phys.} {\bfseries 130} (2008) 129--168},
  \href{http://arxiv.org/abs/0706.3384}{{\ttfamily arXiv:0706.3384 [hep-th]}}.

\bibitem{PAM_doi:10.1007/JHEP03(2015)163}
F.~M. Haehl and M.~Rangamani, ``{Permutation orbifolds and holography},''
  \href{http://dx.doi.org/10.1007/JHEP03(2015)163}{{\em {Journal of High Energy
  Physics}} {\bfseries 2015} no.~3, (2015) 1--45}.

\bibitem{Fitzpatrick:2014vua}
A.~L. Fitzpatrick, J.~Kaplan, and M.~T. Walters, ``{Universality of
  Long-Distance AdS Physics from the CFT Bootstrap},''
  \href{http://dx.doi.org/10.1007/JHEP08(2014)145}{{\em JHEP} {\bfseries 08}
  (2014) 145}, \href{http://arxiv.org/abs/1403.6829}{{\ttfamily arXiv:1403.6829
  [hep-th]}}.

\bibitem{Fitzpatrick:2015zha}
A.~L. Fitzpatrick, J.~Kaplan, and M.~T. Walters, ``{Virasoro Conformal Blocks
  and Thermality from Classical Background Fields},''
  \href{http://dx.doi.org/10.1007/JHEP11(2015)200}{{\em {Journal of High Energy
  Physics}} no.~11, (2015) 1--32},
  \href{http://arxiv.org/abs/1501.05315}{{\ttfamily arXiv:1501.05315
  [hep-th]}}.

\bibitem{Asplund:2014coa}
C.~T. Asplund, A.~Bernamonti, F.~Galli, and T.~Hartman, ``{Holographic
  Entanglement Entropy from 2d CFT: Heavy States and Local Quenches},''
  \href{http://dx.doi.org/10.1007/JHEP02(2015)171}{{\em JHEP} {\bfseries 02}
  (2015) 171}, \href{http://arxiv.org/abs/1410.1392}{{\ttfamily arXiv:1410.1392
  [hep-th]}}.

\bibitem{Roberts:2014ifa}
D.~A. Roberts and D.~Stanford, ``{Two-dimensional conformal field theory and
  the butterfly effect},''
  \href{http://dx.doi.org/10.1103/PhysRevLett.115.131603}{{\em Phys. Rev.
  Lett.} {\bfseries 115} no.~13, (2015) 131603},
  \href{http://arxiv.org/abs/1412.5123}{{\ttfamily arXiv:1412.5123 [hep-th]}}.

\bibitem{hep-th/9901012v2}
G.~T. Horowitz and N.~Itzhaki, ``{Black Holes, Shock Waves, and Causality in
  the AdS/CFT Correspondence},''
  \href{http://dx.doi.org/10.1088/1126-6708/1999/02/010}{{\em JHEP} {\bfseries
  9902} (Jan., 1999) 010},
  \href{http://arxiv.org/abs/hep-th/9901012v2}{{\ttfamily
  arXiv:hep-th/9901012v2 [hep-th]}}.

\bibitem{Nozaki:2013wia}
M.~Nozaki, T.~Numasawa, and T.~Takayanagi, ``{Holographic Local Quenches and
  Entanglement Density},''
  \href{http://dx.doi.org/10.1007/JHEP05(2013)080}{{\em JHEP} {\bfseries 05}
  (2013) 080}, \href{http://arxiv.org/abs/1302.5703}{{\ttfamily arXiv:1302.5703
  [hep-th]}}.

\bibitem{1102.4352v2}
J.~V. Rocha and V.~Cardoso, ``{Gravitational perturbation of the BTZ black hole
  induced by test particles and weak cosmic censorship in AdS spacetime},''
  \href{http://dx.doi.org/10.1103/PhysRevD.83.104037}{{\em Phys.Rev.D}
  {\bfseries 83} (May, 2011) 104037},
  \href{http://arxiv.org/abs/1102.4352v2}{{\ttfamily arXiv:1102.4352v2
  [gr-qc]}}.

\bibitem{Maldacena:2016hyu}
J.~Maldacena and D.~Stanford, ``{Remarks on the Sachdev-Ye-Kitaev model},''
  \href{http://dx.doi.org/10.1103/PhysRevD.94.106002}{{\em Phys. Rev.}
  {\bfseries D94} no.~10, (2016) 106002},
  \href{http://arxiv.org/abs/1604.07818}{{\ttfamily arXiv:1604.07818
  [hep-th]}}.

\end{thebibliography}\endgroup

\end{document}